# Integrating earth observation data into the tri-environmental evaluation of the economic cost of natural disasters: a case study of 2025 LA wildfire

Zongrong Li[a†], Haiyang Li[a†], Yifan Yang[b], Siqin Wang[a*], Yingxin Zhu[a]

[a] Spatial Sciences Institute, University of Southern California, Los Angeles, United States

[b] Department of Geography, Texas A&M University, College Station, United States

Z.L.: zongrong@usc.edu; H.L.: lihaiyan@usc.edu ; Y.Y.: yyang295@tamu.edu S.W.: siqinwan@usc.edu; Y.Z.: yzhu7819@usc.edu

[†] Authors contributed equally to this work.

**Corresponding author:**

Siqin Wang, Spatial Sciences Institute, University of Southern California, Los Angeles, US. siqinwan@usc.edu

**Authors' contributions**

Zongrong Li: Conceptualization, Data curation, Formal analysis, Investigation, Methodology, Software, Validation, Writing – original draft

Haiyang Li: Methodology, Software, Validation, Formal analysis, Investigation, Visualization

Yifan Yang: Conceptualization, Methodology, Validation, Investigation, Data Curation, Writing - Review & Editing

Siqin Wang: Conceptualization, Validation, Investigation, Resources, Writing - Review & Editing, Supervision, Project administration, Funding acquisition

Yingxin Zhu: Software, Investigation, Writing – original draft

**Statements and Declarations**

Competing Interests

The authors declare that they have no competing financial or non-financial interests directly or indirectly related to the work submitted for publication.

# Integrating earth observation data into the tri-environmental evaluation of the economic cost of natural disasters: a case study of 2025 LA wildfire

**Abstract:**

Wildfires in urbanized regions, particularly within the wildland–urban interface, have significantly intensified in frequency and severity, driven by rapid urban expansion and climate change. This study aims to provide a comprehensive, fine-grained evaluation of the recent 2025 Los Angeles wildfire's impacts, through a multi-source, tri-environmental framework in the social, built and natural environmental dimensions. This study employed a spatiotemporal wildfire impact assessment method based on daily satellite fire detections from the Visible Infrared Imaging Radiometer Suite (VIIRS), infrastructure data from OpenStreetMap, and high-resolution dasymetric population modeling to capture the dynamic progression of wildfire events in two distinct Los Angeles County regions, Eaton and Palisades, which occurred in January 2025. The modelling result estimated that the total direct economic losses reached approximately $4.86 billion with the highest single-day losses recorded on January 8 in both districts. Population exposure reached a daily maximum of 4,342 residents in Eaton and 3,926 residents in Palisades. Our modelling results highlight early, severe ecological and infrastructural damage in Palisades, as well as delayed, intense social and economic disruptions in Eaton. This tri-environmental framework underscores the necessity for tailored, equitable wildfire management strategies, enabling more effective emergency responses, targeted urban planning, and community resilience enhancement. Our study contributes a highly replicable tri-environmental framework for evaluating the natural, built and social environmental costs of natural disasters, which can be applied to future risk profiling, hazard mitigation, and environmental management in the era of climate change.

## 1. Introduction

Wildfire has become an increasingly urban problem in the western United States as metropolitan growth pushes farther into flammable landscapes: in California alone, the wildland–urban interface (WUI) expanded by more than 30% between 1990 and 2020, situating millions of residents and billions of dollars in assets in places where severe fire is no longer an anomaly but an annual expectation (Radeloff et al., 2018; Chow et al., 2022). Recent events underscore the stakes: the 2018 Camp Fire leveled the town of Paradise; the 2021 Dixie Fire scorched nearly one million acres; and a series of fast-moving blazes in early 2025 threatened multiple neighborhoods in Los Angeles County (Matt, 2024; Tutella, 2021). Beyond the immediate loss of life and property, such fires interrupt economic activity, degrade air quality, and exacerbate long-term ecological stressors such as invasive species spread and post-fire soil erosion (João Gonçalves et al., 2025). Yet most wildfire impact assessments tend to treat these consequences in isolation—mapping burn scars one season at a time, tabulating insurance claims, or profiling social vulnerability—without analyzing how the natural, built and social environment interact dynamically during the course of a single fire event. This research addresses this gap by contributing a fine-grained, tri-environmental framework that captures the full spatial and temporal complexity of the wildfire impact.

A substantial body of research has advanced our understanding of wildfire behavior and consequences, but often through discipline-specific lenses. Ecological studies focus on vegetation loss and land degradation using spectral indices such as differenced normalized burn ratio (dNBR) (Ibrahim et al., 2024), while infrastructure assessments examine damage to roads, buildings, and utilities. Social vulnerability research highlights the disparate capacities of populations to respond to and recover from fire events, often linked to age, income, ethnicity, and housing conditions (Moore et al., 2023). However, these studies typically operate at coarse spatial or temporal scales or rely on static data snapshots that fail to capture the evolving nature of fire exposure. A few recent researchers explored integrative approaches that combine environmental, built, and social variables particularly in the context of floods and heatwaves (Wang et al., 2023; Raymond et al., 2020). In wildfire research, such comprehensive and real-time evaluations across environmental, built, and social dimensions are much needed to support timely response strategies for complex events such as the 2025 LA Wildfire.

Advances in earth-observation technologies and open geospatial data now offer an opportunity to close this integration gap. Near-real-time satellite sensors such as Moderate Resolution Imaging Spectroradiometer (MODIS) and Visible Infrared Imaging Radiometer Suite (VIIRS) provide frequent thermal detections that enable tracking of fire progression at daily intervals. Platforms such as the Fire Information for Resource Management System (FIRMS), OpenStreetMap (OSM), and the National Land Cover Database (NLCD) offer globally accessible layers on active fire points, infrastructure, and land use, respectively (NASA FIRMS, 2025; Liu et al., 2024). Furthermore, dasymetric mapping techniques have made it possible to downscale census populations to grid resolutions as fine as 10–30 meters, redistributing demographic data across heterogeneous land cover types and revealing who lives where with much greater precision than traditional census tracts (Swanwick et al., 2022; Depsky et al., 2022). However, despite the availability of these tools, few studies have developed workflows that simultaneously and dynamically incorporate these multi-source data streams to assess daily wildfire impacts across natural, built and social environments.

To address these knowledge gaps, this study introduces and implements a multi-source, tri-environmental framework for wildfire impact analysis, integrating natural, built, and social environmental evaluation at a fine spatial and temporal resolution (20 meters by 20 meters on the daily basis). The framework is demonstrated through application to two distinct WUI regions in Los Angeles County—Eaton and Palisades—each characterized by contrasting topography, vegetation, settlement patterns, and demographic profiles. The reconstructed daily wildfire perimeters, derived from VIIRS thermal anomalies and refined using CAL FIRE official boundaries, are overlaid with NLCD-based land cover data, OSM-derived built environment features (including road networks and building footprints), Foursquare Points of Interest (POI) data, and a 20-meter resolution dasymetric population grid. This layered spatial analysis enables the disaggregation of wildfire impacts by environment type and population vulnerability at a daily basis.

Through the tri-environmental evaluation, this paper makes three key contributions. First, it presents a reproducible framework for daily-scale fire boundary reconstruction and impact quantification using publicly available data. Second, it operationalizes a tri-environmental framework that captures the interdependencies among ecological exposure, infrastructure disruption, and social vulnerability, which are too often treated separately in disaster research. Third, it demonstrates the value of high-resolution, intra-urban comparative analysis through a case study of Eaton and Palisades, offering practical insights into how wildfire risk manifests differently across neighborhoods. Ultimately, the findings underscore the need for integrated, spatially explicit, and equity-aware wildfire management strategies that align emergency response, urban planning, and community resilience efforts.

## 2. Literature Review

### *2.1 Wildfire Impacts in Wildland–Urban Interface (WUI) Zones*

Wildfires in the WUI, where urban structures intersect with wildland vegetation, are becoming increasingly devastating. In California, rapid WUI expansion into flammable landscapes has sharply elevated risk (Kumar et al., 2025). From 2000 to 2018, about 54,000 U.S. buildings were lost to wildfires, with 86–97% located in the WUI (Caggiano et al., 2020). California's WUI growth and climate change have together increased extreme fire conditions 4.1-fold since 1990 (Kumar et al., 2025). Most major fires now originate within 1 km of WUI areas, underscoring wildfire as a predominantly WUI-driven hazard (Radeloff et al., 2018).

Damage patterns in WUI fires are multifaceted. Structural losses often occur in clusters, destroying entire neighborhoods, as seen in the 2018 Camp Fire and 2025 Los Angeles fires. Beyond property loss, wildfires cause severe ecological degradation, including habitat fragmentation, biodiversity loss, erosion, and nutrient depletion (Troy et al., 2022), often worsened by human activities like fire suppression and invasive species. Socially vulnerable populations face disproportionate wildfire exposure, which rose significantly from 2000 to 2021, especially among low-income, elderly, and mobility-impaired residents in California (Modaresi Rad et al., 2023; Masri et al., 2021).. However, most of the previous WUI studies focused on a single perspective of the built environment (home ignition risks, building materials, defensible space), natural environment (fire regimes, ecological impacts), or social aspects (vulnerability mapping, evacuation logistics)—limiting the comprehensive understanding of wildfire impacts within WUI zones.

*2.2 Advances in Daily Wildfire Monitoring and Mapping*

Recent technological advances facilitate daily-scale wildfire monitoring, improving management and research capabilities. Historically, burned areas were assessed post-event, but current satellite technologies such as NASA's MODIS and the VIIRS now offer near real-time data at resolutions of approximately 1 km and 375 m, respectively. NASA's FIRMS also provides continuous global fire hotspot data, significantly enhancing wildfire situational awareness (NASA FIRMS, 2025).

Advanced techniques now convert satellite fire detections into actionable insights. Algorithms cluster thermal pixels into dynamic perimeters, enabling real-time updates—some systems generate fire boundaries every 12 hours, improving fire behavior monitoring (Chen et al., 2022). This approach offers faster insights than post-event imagery, aiding daily fire tracking and model evaluation. Spatial tools like Kernel Density Estimation (KDE) reveal ignition hotspots and shifting fire patterns, both historically and in real time. Real-time KDE supports emergency response by highlighting active fire zones. Integrating fire perimeters with demographic and infrastructure data via tools ReadyMapper, helps managers identify threats to vulnerable populations and critical services, enabling timely, targeted evacuation and resource deployment (Schroeder et al., 2023).

However, despite available technology, literature gaps persist. Most wildfire studies remain retrospective, lacking full utilization of daily satellite data granularity. There is limited published research operationalizing daily tracking in impact assessments or systematically integrating fire behavior data with socio-demographic information. Addressing these gaps could significantly enhance wildfire impact assessments.

*2.3 Tri-Environmental Framework for Disaster Research*

The tri-environmental framework, encompassing the natural, built, and social environments, offers a holistic approach to disaster analysis, recognizing disasters as products of interconnected systems rather than isolated phenomena Wang et al. (2023). Originating from social-ecological theories, this framework emphasizes comprehensive disaster assessments, incorporating interactions among physical environments, infrastructure, and human communities.

Recent studies applying the tri-environmental framework illustrate its effectiveness. For instance, Wang et al. (2023) proposed this framework in Australian heatwave risk assessments, integrating environmental hazards, urban characteristics, and population vulnerabilities. Similarly, Aquilino et al. (2021) enhanced urban sustainability indicators by combining fine-scale demographic, land use, and infrastructure data to improve population mapping accuracy. Their methodology allowed precise evaluations of who might be impacted by environmental hazards, demonstrating the value of integrated spatial analyses in urban planning and risk assessment. Despite its potential, the tri-environmental framework has limited application in high-resolution wildfire-specific studies. A recent review highlighted a significant gap in integrating wildfire's natural, built, and social dimensions into cohesive analyses (Li et al., 2024). Few studies have combined detailed environmental data with demographic analyses for comprehensive wildfire impact assessment at city or fire-event scales. The tri-environmental concept remains largely theoretical in wildfire contexts, with existing urban resilience models often focused on floods or heat waves rather than wildfires.

This study seeks to address this research gap by implementing a tri-environmental framework specifically designed for wildfire analysis in Los Angeles. By moving beyond traditional compartmentalized approaches, the framework enables a nuanced understanding of how interactions among vegetation (natural environment), infrastructure configurations (built environment), and community demographics (social environment) shape wildfire impacts. Such an integrated perspective enhances the precision of risk assessments and informs the development of targeted mitigation strategies to support resilience in wildfire-prone urban environments.

## 3. Study area and data

### 3.1 Study area

Eaton and Palisades (Fig. 1) are two wildfire-prone regions located in Los Angeles County, California. Both represent distinct examples of WUI environments, where human development intersects with fire-adapted natural landscapes. These areas are characterized by complex terrain, diverse vegetation assemblages, and a high frequency of wildfire exposure (Radeloff et al., 2018).

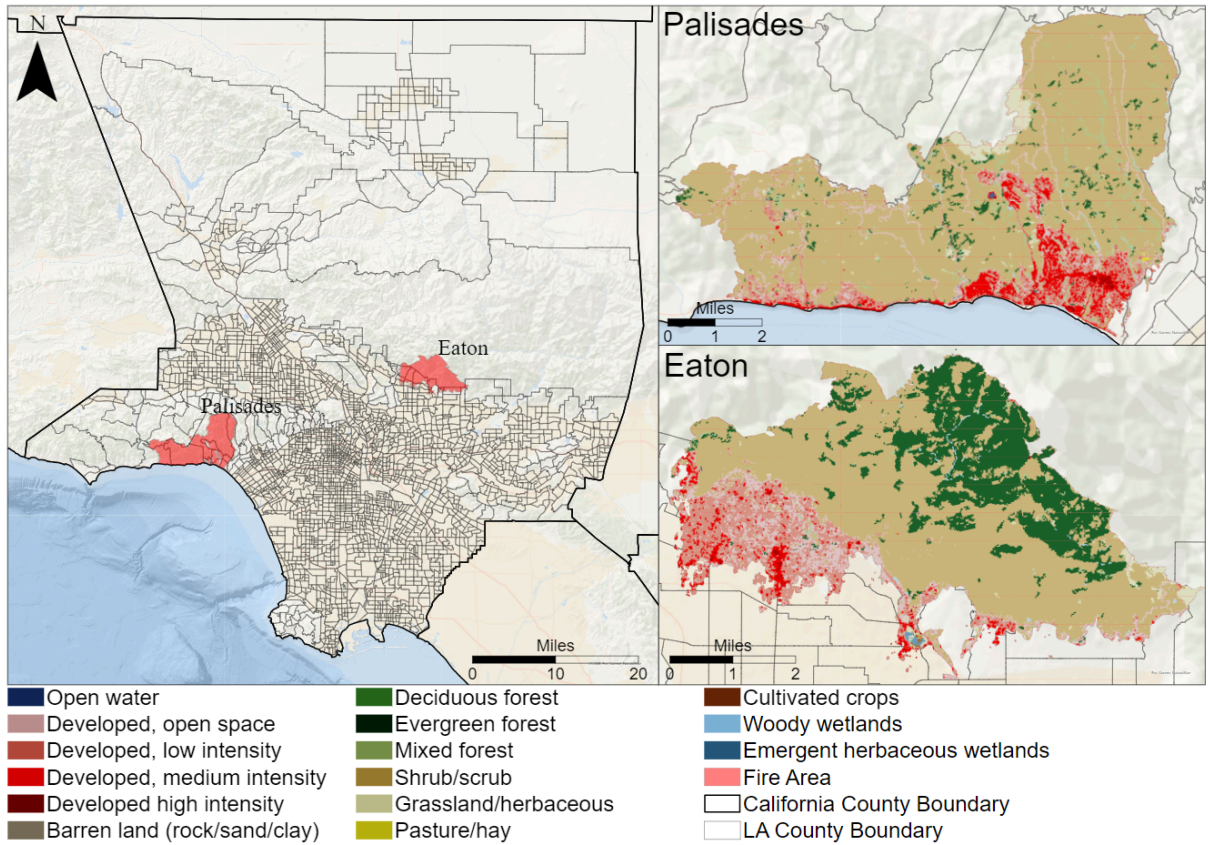

Figure 1. Wildfire-Affected Areas in Palisades and Eaton, Los Angeles

Note: The left panel shows the locations of Palisades and Eaton in Los Angeles County. The right panels display enlarged land cover and burned areas in each fire zone.

Eaton (Fig. 1, bottom right), located along the foothills of the San Gabriel Mountains, is characterized by steep canyons, dense evergreen forests, and extensive chaparral. Medium- to low-density residential development interspersed with wildland vegetation forms a classic intermix WUI. Wildfire vulnerability is heightened by Santa Ana winds that funnel through canyons, accelerating downslope fire spread, and by limited evacuation infrastructure (Abatzoglou et al., 2023). The spatial proximity of flammable vegetation and housing complicates suppression and emergency response. Palisades (Fig. 1, top right) occupies the coastal slopes of the Santa Monica Mountains, where a Mediterranean climate with prolonged dry summers and offshore winds, creates conditions favorable for ignition and rapid fire spread. The area comprises a mosaic of residential zones, wildland parks (e.g., Topanga State Park), and undeveloped shrublands. Steep terrain and dense housing adjacent to vegetated hillsides form a continuous fuel matrix across the WUI (Syphard et al., 2012). Both regions face increasing wildfire frequency and intensity due to climate-driven aridification, fuel accumulation, and the proliferation of invasive grasses, highly flammable surface fuels that exacerbate fire behavior amid continued urban expansion (Balch et al., 2022; Modaresi Rad et al., 2023).

*3.2 Data collection and manipulation*

To assess the natural, built and social environments of the 2025 Los Angeles wildfire, we constructed a comprehensive spatial dataset library that integrates fire boundaries, thermal activity, land cover, demographic, and POI information. These datasets were selected based on spatial resolution, update frequency, and thematic relevance to wildfire impact analysis (Table 1).

Table 1. Sources of Datasets Included in the Spatial Data Layers Library

| Attribute | Los Angeles fire boundary | VIIRS NOAA-21 | OSM Data | Land Cover Data | Census Data | Foursquare Open Source Places |
|---|---|---|---|---|---|---|
| Category | Fire Information | Daily wildfire spread | OpenStreetMap | Land Cover | Demographics | POI Data |
| Source | CAL FIRE | NASA | OpenStreetMap | USGS/NLCD | US Census Bureau | Foursquare |
| Format | Vector | Raster | Point | Raster | Vector | Point |
| Spatial Resolution | N/A | 375m | N/A | 30 m | County level | Global Level |
| Temporal Resolution | N/A | Daily | N/A | Annual | Decennial | Annual |
| Time | 07/01/2025–13/01/2025 | 07/01/2025–13/01/2025 | 2024 | 2021 | 2020 | 2024 |

Note: CAL FIRE: California Department of Forestry and Fire Protection; VIIRS: Visible Infrared Imaging Radiometer Suite; NOAA: National Oceanic and Atmospheric Administration; OSM:

OpenStreetMap; NLCD: National Land Cover Database; USGS: United States Geological Survey.

**Fire Boundary and Thermal Activity.** The wildfire perimeter dataset from the California Department of Forestry and Fire Protection (CAL FIRE) delineates the official burn extent from July 1 to January 13, 2025 (CAL FIRE, 2025). To capture temporal dynamics, this study uses daily thermal anomaly data from the VIIRS sensor aboard NOAA-21. VIIRS provides 375-meter resolution imagery based on mid-infrared signatures and supports near-real-time detection of active fire fronts (Schroeder et al., 2014).

**Land Cover and Built Environment.** For environmental context and dasymetric modeling (detailed in the later subsection), the study uses the 2021 National Land Cover Database (NLCD), produced by the U.S. Geological Survey. This 30 m resolution raster dataset classifies land cover types nationwide and informs both vegetation-related fire behavior analysis and population weight assignment in dasymetric mapping (Dewitz, 2023). Built environment data are obtained from OpenStreetMap (OSM, 2024), an open-source, point-based dataset that includes roads, buildings, and land use features. OSM is commonly applied in disaster research for mapping infrastructure and accessibility (Haklay & Weber, 2008).

**Demographics and Social Indicators.** Demographic data are sourced from the U.S. Census Bureau (2020) at the county level, providing baseline population estimates for dasymetric downscaling. To enhance spatial resolution, the study incorporates Foursquare Open Source Places data (2024), which includes global points of interest (POIs) across residential, commercial, and public service categories. These POIs help identify human activity centers and inform built environment impact assessments (Foursquare Labs, 2024).

## 4. Methodology

This study adopts a multi-step geospatial methodology to assess wildfire impacts across natural, built, and social environments in two WUI zones in Los Angeles County. The framework integrates multiple data sources—thermal satellite imagery, official fire boundaries, land cover classifications, infrastructure layers, and dasymetrically mapped population data—into a spatially explicit and temporally dynamic assessment. The workflow consists of three core components: (1) delineation of daily wildfire extents, (2) downscaling of population density using dasymetric mapping, and (3) integrated impact evaluation through a tri-environmental lens, as illustrated in Figure 2.

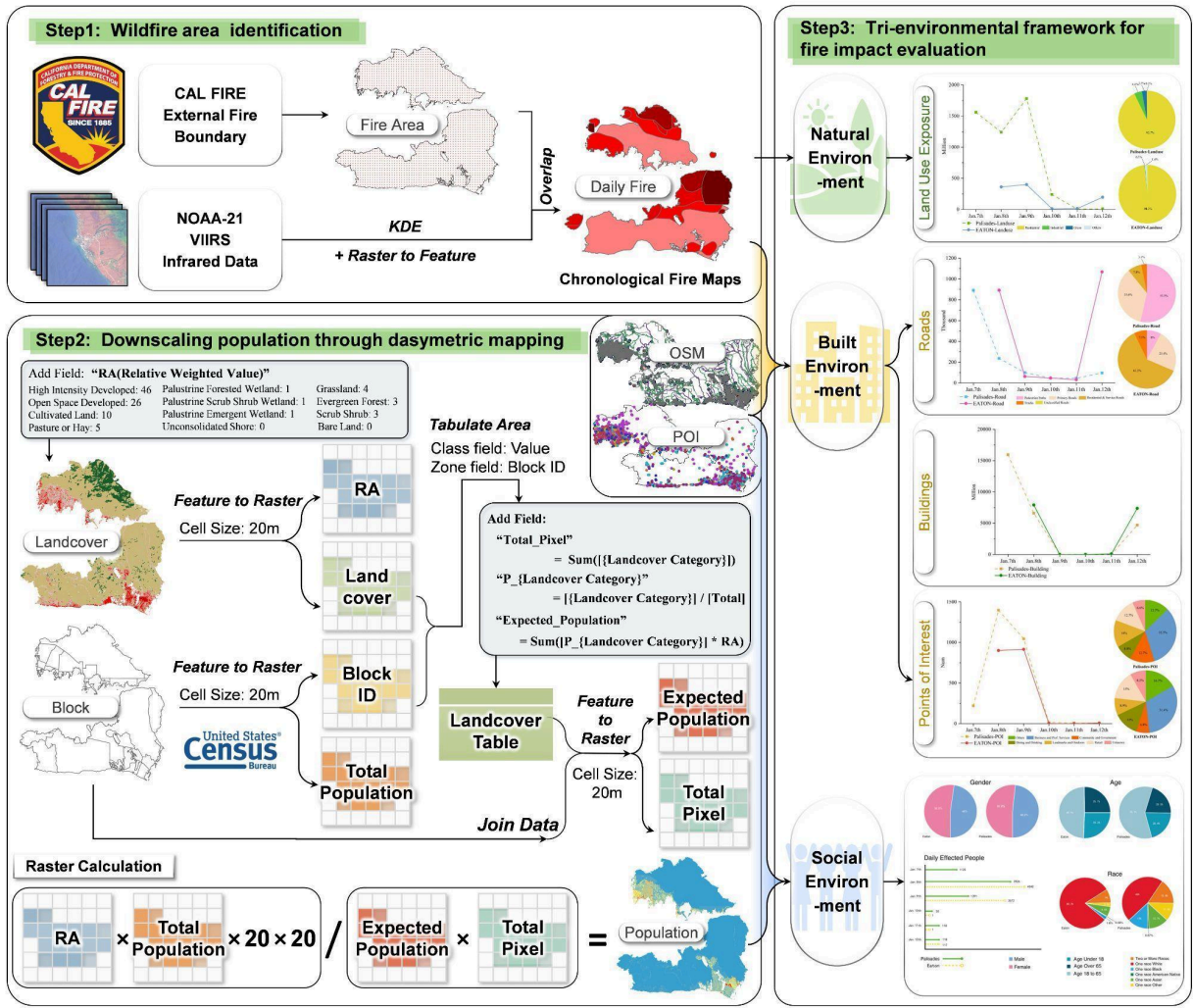

Figure 2. Workflow of the Tri-Environmental Wildfire Impact Assessment Framework

## 4.1 Wildfire area identification

To accurately assess the spatial extent and temporal progression of the 2025 Los Angeles wildfire, we implemented a hybrid framework that integrates official fire perimeter data with satellite-derived thermal observations (Fig. 2, Step 1). The static fire boundary was obtained from CAL FIRE, delineating the overall extent of the burned area from July 1 to January 13,

2025. While this vector dataset provided a reliable outline of cumulative fire impact, it lacked the temporal granularity needed to track daily fire dynamics.

To capture day-by-day fire activity within this perimeter, we incorporated thermal anomaly data from the VIIRS aboard the NOAA-21 satellite. This dataset delivered 375-meter resolution nighttime imagery, enabling high-frequency detection of active fire fronts based on mid-infrared radiative signals. We applied KDE to the VIIRS point detections, generating a continuous surface that highlights zones of concentrated heat activity and frequent ignition. The resulting raster was thresholded to distinguish burn-affected areas and subsequently converted into vector polygon format, producing a sequence of daily fire boundaries. By intersecting these daily polygons with the CAL FIRE perimeter, we generated refined, temporally disaggregated fire maps that integrated both satellite-detected hotspots and officially reported perimeters.

## *4.2 Downscaling population through dasymetric mapping*

Conventional gridded population datasets such as LandScan (2022) and WorldPop (2021) offer valuable population estimates, but their spatial resolutions (1 km and 100–1000 m, respectively) are insufficient for detailed wildfire impact analysis at the neighborhood or WUI scale. To overcome this limitation, we employed a dasymetric mapping approach that downscaled population data to a 20 m × 20 m resolution, allowing more precise identification of exposed populations.

Dasymetric mapping is a spatial redistribution method that disaggregates census-level population data using land cover and ancillary variables to reflect the physical plausibility of habitation

(Mennis, 2003; Leyk et al., 2019). In our study, we incorporated 2020 U.S. Census block-level data and the 2021 NLCD to derive population density surfaces that account for heterogeneity in land use across Eaton and Palisades.

The dasymetric mapping process (Fig. 2, Step 2) involved a multi-step workflow to redistribute population data based on land use suitability. First, we assigned Relative-weight Values (RA) to each land cover class according to its potential to host human populations. Developed areas—particularly those with high residential intensity—received higher weights, while natural features such as forests, wetlands, and open water were assigned minimal or zero values (Table 2). These RA values were used to generate the Population Allocation Factor Raster, as expressed in Equation (1):

$$RA = f(LandUseType) \tag{1}$$

Next, we used the Tabulate Area tool to calculate the total number of pixels associated with each land cover class in each census block, as shown in Equation (2):

$$TotalPixel = \sum (PixelCount) \tag{2}$$

These values allowed us to compute the proportion of each land cover type within a block and estimated the expected population for each class accordingly. Finally, we normalized and adjusted population values using the Adjusted Population Formula in Equation (3):

$$AdjustedPopulation = (RA \times Pop \times 400) / (TotalPixel \times ExpectedPopulation) \quad (3)$$

Here, Pop was the total census population for a block, RA was the relative weight for the given land cover class, and 400 represents the area in square meters of each raster cell (20 m × 20 m). This adjustment step ensured that the population was distributed in accordance with the physical land suitability and spatial heterogeneity across the WUI landscape.

Table 2. Relative Weight Values Assigned to Land Cover Types for Dasymetric Mapping

| Land Cover Type | Relative Weighted Value (RA) |
|---|---|
| Open water | 0 |
| Developed, open space | 26 |
| Developed, low intensity | 10 |
| Developed, medium intensity | 15 |
| Developed, high intensity | 46 |
| Barren land | 0 |
| Deciduous forest | 3 |
| Evergreen forest | 3 |
| Mixed forest | 4 |
| Shrub/scrub | 3 |
| Grassland/herbaceous | 4 |
| Pasture/hay | 5 |
| Cultivated crops | 10 |
| Woody wetlands | 1 |
| Emergent herbaceous wetlands | 1 |

Note: Relative-weighted Values (RA) reflected the suitability of each land cover type for population distribution, with higher weights for developed areas and lower or zero weights for natural or uninhabited land.

### *4.3 Tri-environmental framework for fire impact evaluation*

To comprehensively evaluate the multidimensional effects of the 2025 Los Angeles wildfire, we implemented a tri-environmental framework that assessed impacts across three interconnected domains: the natural environment, the built environment, and the social environment (Fig. 2, Step 3). This integrative framework enabled a holistic understanding of wildfire consequences by combining geospatial, infrastructural, and demographic datasets within the defined fire-affected zones.

**Natural and Built Environment Assessment.** We began by overlaying the refined wildfire polygons (derived in Section 4.1) with vector-based datasets from OpenStreetMap (OSM, 2024), which provided detailed geospatial representations of building footprints, road networks, and landscape features such as parks and vegetated areas. This spatial intersection allowed us to identify infrastructure elements and ecological land cover types that fell within the fire perimeter, offering direct evidence of physical impact. To assess disruptions to community-serving infrastructure, we also incorporated Foursquare Open Source Places (2024), which catalogued a wide array of POIs across Los Angeles, including hospitals, schools, retail businesses, and restaurants. Intersecting POI locations with the wildfire boundary enabled us to quantify functional losses and evaluate the exposure of critical urban amenities to wildfire threats.

**Social Environment Assessment**. To analyze the fire's social impact, we intersected the daily wildfire footprint with the high-resolution population density surface generated through dasymetric mapping (Section 4.2). This overlay allowed for precise estimation of the number of residents exposed to fire hazards on a per-pixel basis, rather than relying on coarse administrative units. We further enriched this spatial analysis by integrating 2020 U.S. Census demographic data, which enabled disaggregation of affected populations by age group, gender, and income level at the census tract scale. Through this process, we identified vulnerable

subpopulations—such as children, elderly individuals, and low-income households—residing within or adjacent to the fire-affected zones. By linking demographic profiles with spatial exposure data, this framework highlighted patterns of social vulnerability that were critical for informing targeted evacuation, relief distribution, and long-term recovery planning.

**Cross-Comparative Economic Loss Estimates with Other Assessments.** To evaluate the validity and applicability of our economic loss estimation framework, we incorporated a cross-comparative design with two regional-scale assessments of the 2025 Los Angeles wildfire: the UCLA Anderson Forecast and the Los Angeles County Economic Development Corporation (LAEDC) report. This comparative structure enabled us to assess the internal consistency of our estimates and explore how methodological differences—such as scale, input data, and attribution logic—could influence the interpretation of wildfire-related economic consequences. Although the focus of this study remained on localized, asset-level losses, aligning our estimates with broader benchmarks provided an important lens for understanding both the strengths and the boundaries of spatially disaggregated disaster impact models.

## 5. Results

### 5.1 Wildfire Progression and Spatial Dynamics

The spatiotemporal dynamics of wildfire progression in both Eaton and Palisades are illustrated in Figure 3, which displays daily burned areas and aggregated ignition points from January 7 to January 12, 2025. This visual record, based on combined CAL FIRE boundaries and VIIRS thermal detections, captures the evolving nature of fire behavior across two distinct topographic contexts within Los Angeles County.

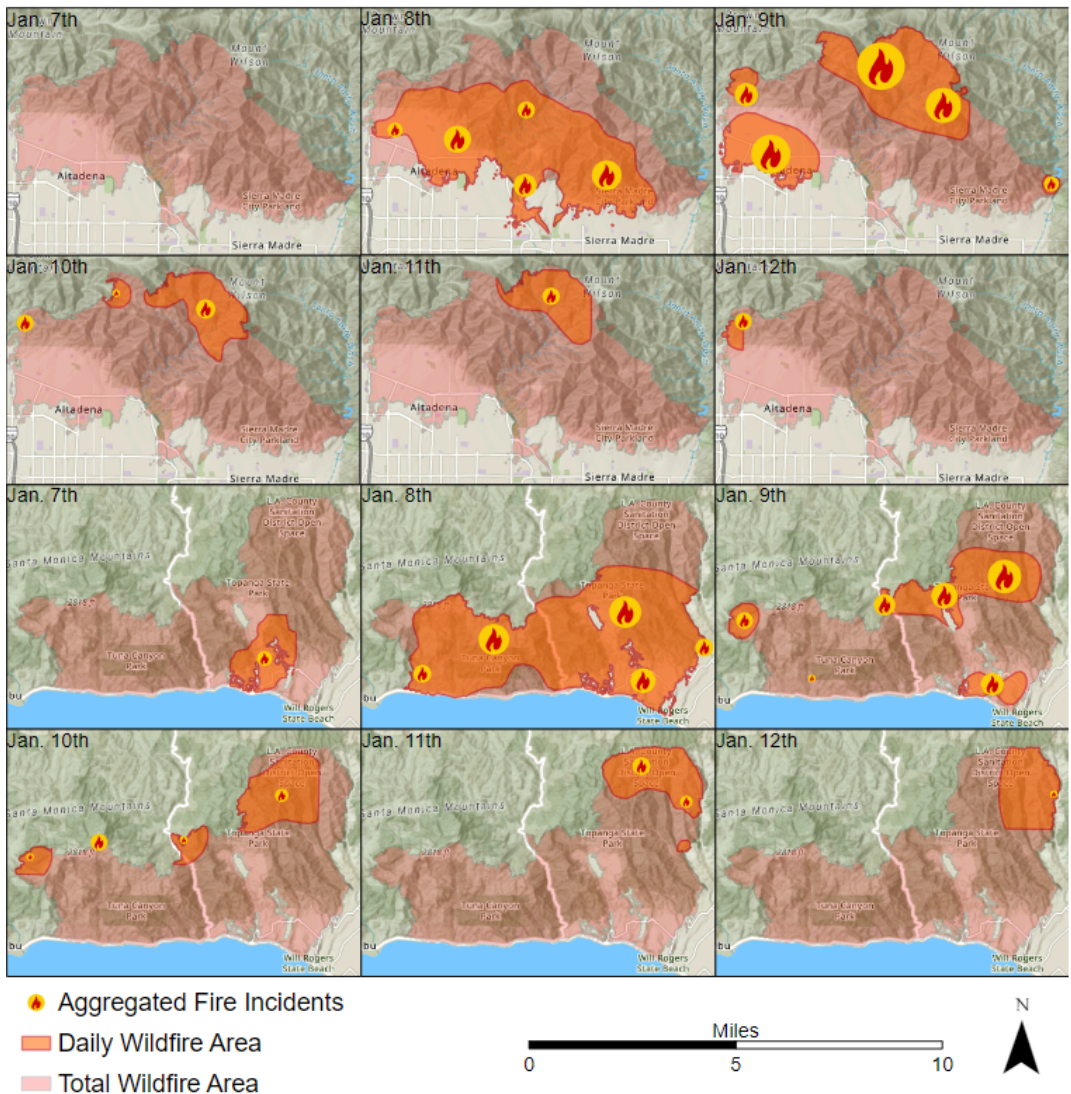

Figure 3. Daily Progression of Wildfire Spread in Eaton and Palisades (January 7–12, 2025)

Note: The figure shows daily fire spread in Eaton (top six maps) and Palisades (bottom six maps) using VIIRS thermal detections and CAL FIRE perimeters. Orange polygons indicate new burns; pink shading shows cumulative extent; flame icons mark daily hotspots.

In Eaton, wildfire activity initiates on January 7 with localized ignitions on north-facing slopes of the San Gabriel Mountains, northeast of Mount Wilson. These early burns occur in steep, densely vegetated terrain conducive to fire spread. By January 8, the fire advances southeast along the Altadena–Chaney Trail corridor toward the urban fringe. On January 9, it undergoes substantial expansion, forming a continuous burn front from eastern Altadena toward Sierra Madre. A new ignition on January 10 near Millard Canyon, possibly caused by ember spotting, adds spatial complexity. Activity becomes fragmented by January 11, with scattered clusters moving southwest and upslope. On January 12, only low-intensity residual burning persists near Las Flores Canyon. The cumulative burn scar spans rugged forested slopes and WUI edges, highlighting suppression challenges in topographically complex terrain.

In contrast, wildfire activity in the Palisades region is first observed on January 7 within Topanga State Park, near Entrada Road and Trippet Ranch, areas known for steep ridgelines and flammable chaparral. On January 8, it spreads bidirectionally, east toward Will Rogers State Park and west into Santa Ynez Canyon, driven by local wind and terrain dynamics. Peak activity occurs on January 9, forming a continuous ignition belt across Tuna Canyon and Los Liones Trail, threatening ridgeline residences. Between January 10–11, the fire progresses along Temescal Ridge and Rustic Canyon, with irregular daily spread indicative of intense flanking and potential crown fire behavior. By January 12, only minor hotspots remain in Rivas and Hondo Canyons. The final burn footprint reveals dense ignition patterns across natural and developed areas, underscoring high WUI exposure and ecological vulnerability.

*5.2 Impact analysis for the natural & built environment*

Wildfires in Eaton and Palisades have markedly different impacts across both natural and built environments. Through the integration of temporal fire boundary data and spatial overlays with land use, infrastructure, and POI datasets, we evaluate daily patterns and cumulative consequences on critical landscape and human systems. Detailed numerical results of exposure and economic loss estimates can be found in Appendix A.

*5.2.1 Natural Environment: Land Use Exposure*

Wildfire exposure to natural and residential land uses differs significantly between the two study areas. As illustrated in Figure 4, land-related economic losses peak in Palisades on January 9, exceeding 17 million USD, while Eaton reaches its highest value (approximately 4 million USD) on the same day. These peaks coincide with the fire's expansion into vegetated hillsides and low-density residential zones.

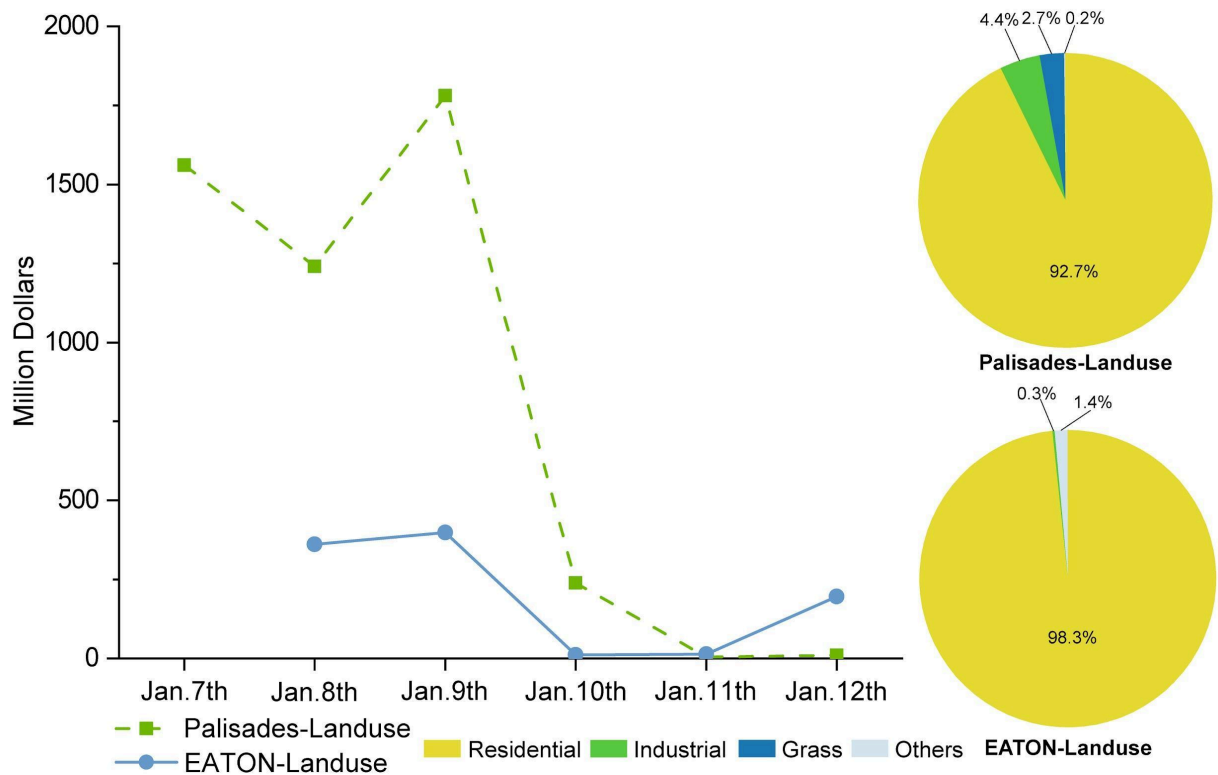

Figure 4. Temporal Changes and Composition of Land Use Affected by Wildfire

Note: The line chart shows daily losses by land use type; pie charts depict total loss composition, with residential areas contributing the largest share in both regions.

The land use composition differs substantially between the two regions. In Eaton, 98.32% of the affected land falls into residential categories, with only 1.42% identified as grass and a minimal 0.26% attributed to industrial or other uses. In contrast, Palisades displays a more diverse composition: 92.73% residential, but with higher proportions of grass (4.43%) and industrial/open land (2.69%). This suggests a greater variety of ecological zones being impacted on Palisades, particularly open hillsides and transitional vegetation belts.

### *5.2.2 Built Environment: Roads*

Wildfire damage to road networks also varies in timing and scale. As shown in Figure 5, Palisades experiences early exposure, with road losses peaking on January 7 at approximately 900,000 USD. This reflects initial fire ignition near recreational and accessible areas. In contrast, Eaton displays delayed impact, with minimal road damage until January 12, when losses surge above 1 million USD, indicating the fire's late incursion into residential transport corridors.

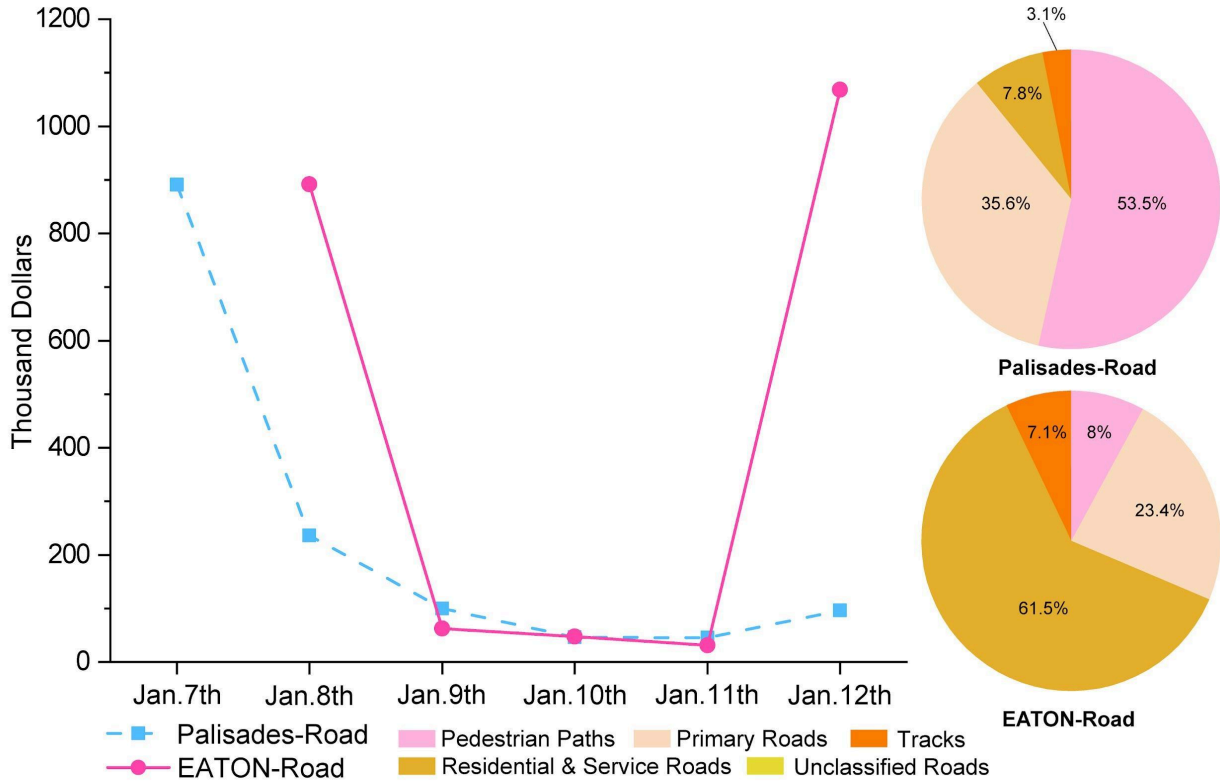

Figure 5. Temporal Changes and Composition of Road Infrastructure Affected by Wildfire

Note: The line chart shows daily road loss estimates; pie charts depict affected road types, with pedestrian paths dominant in Palisades and residential/service roads in Eaton.

Road classification pie charts reveal striking differences: in Eaton, 78.91% of exposed roads are residential and service roads, indicating vulnerability within neighborhoods. Primary roads and unclassified roads contribute 9.15% and 10.21%, respectively. In contrast, Palisades experiences a more diverse distribution: only 42.76% of exposed roads are residential, while 28.45% are primary roads, 6.21% are tracks, and 2.47% are unclassified paths. This suggests a broader impact on access routes and recreational trails in the Palisades region.

*5.2.3 Built Environment: Buildings*

As shown in Figure 6, building-related economic losses are substantial in both regions, with timing and intensity reflecting differences in fire entry into built environments. In Palisades, building damage peaks early, on January 7, with estimated losses exceeding $1.59 billion, corresponding to high-value residential zones near early ignition points. Eaton's building losses lag behind, with a dramatic rise on January 12, reaching over $736 million, as the fire spreads southeast into populated tracts.

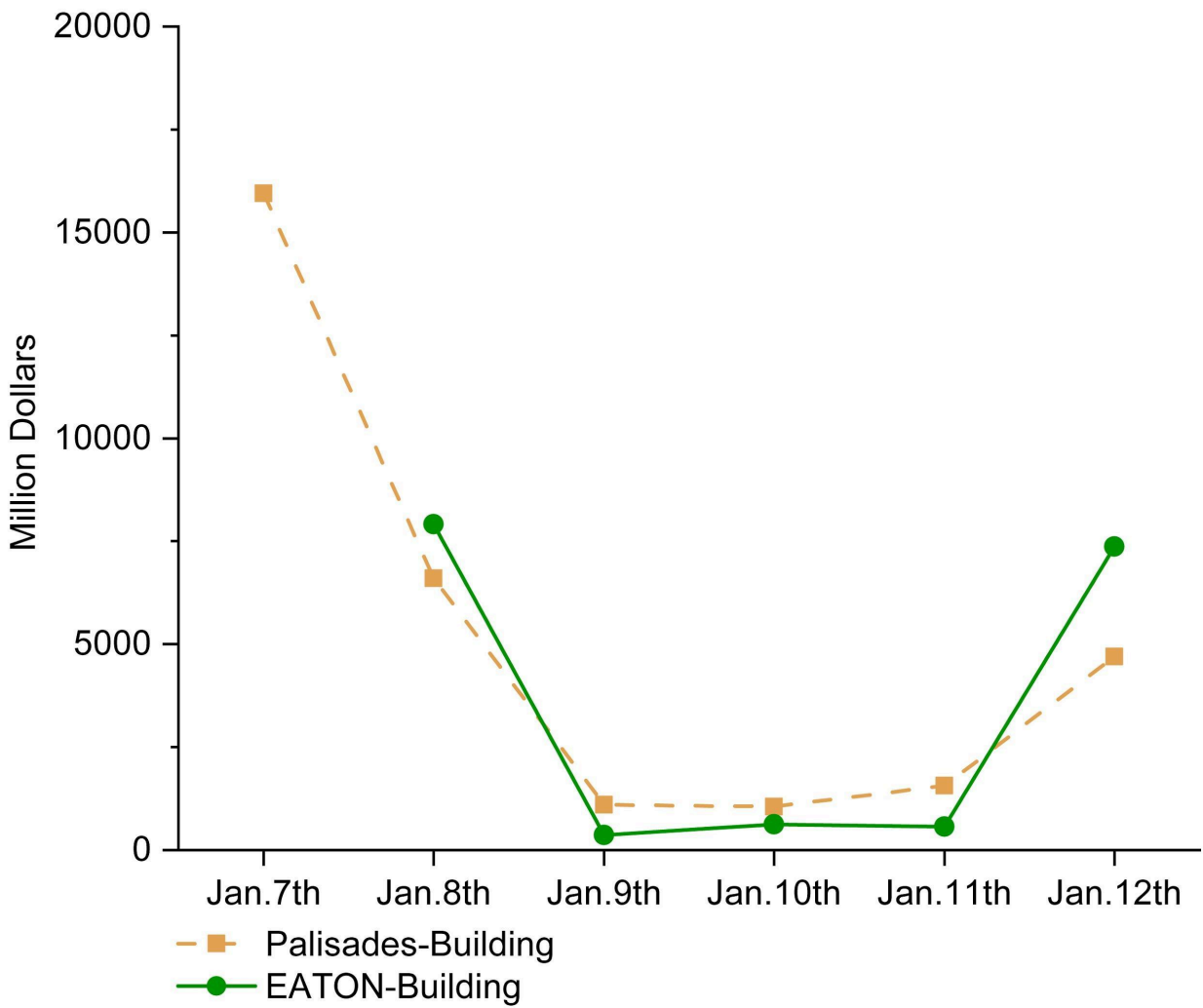

Figure 6. Daily Economic Losses from Fire-Damaged Buildings

Note: The line chart shows daily estimated economic losses from building-related losses in Palisades and Eaton.

Although this dataset does not categorize buildings by type (e.g., residential vs. commercial), the spatial and temporal patterns clearly reflect the fire's encroachment into built-up environments, with early onset in Palisades and lagging but severe impact in Eaton.

### 5.2.4 Built Environment: Points of Interest (POIs)

POIs exposure patterns closely mirror wildfire advancement. As depicted in Figure 7, Eaton records its highest number of affected POIs on January 8 (1,397), while Palisades peaks on January 9 (915). After January 10, both regions show a sharp decline, reflecting either successful containment or the fire's retreat from core urban areas.

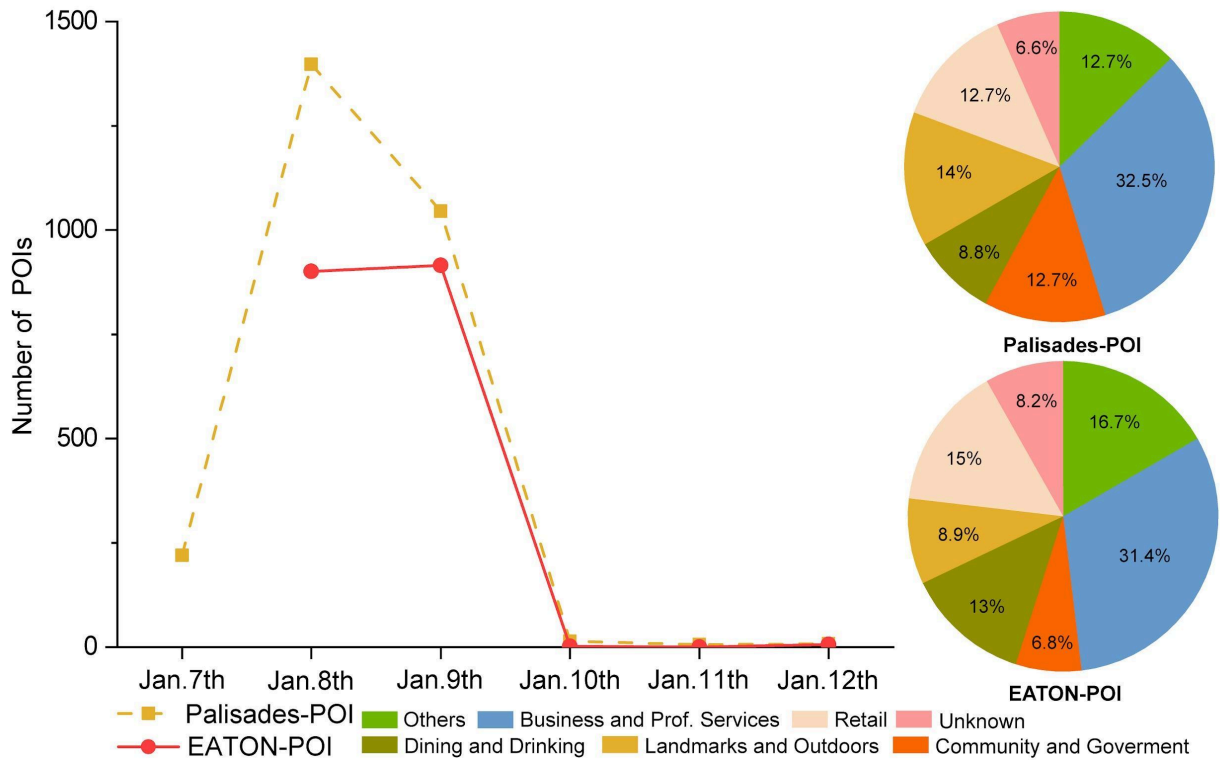

Figure 7. Temporal Change and Type Composition of Fire-Affected POIs

Note: The Line chart shows daily counts of affected Points of Interest (POIs); pie charts display the category breakdown in Palisades and Eaton.

POI type distributions further highlight differences in community impact. In Eaton, the largest affected categories are Business and Professional Services (34.74%), followed by Retail (14.99%), Dining and Drinking (13.55%), and Community and Government services (13.64%). In Palisades, POIs are more evenly distributed, with Business Services (24.08%), Community and Government (12.76%), and Retail (11.46%) making up the majority. This indicates that Eaton's impact is more economically focused, while Palisades affects a broader range of public-facing and mixed-use functions.

*5.3 Impact analysis for the social environment*

To assess the social impacts of the 2025 Los Angeles wildfire, we intersect high-resolution population density layers, generated through dasymetric mapping with daily wildfire perimeters. This approach enables fine-scale analysis of both the temporal progression and demographic composition of exposed populations in Eaton and Palisades—two WUI communities with distinct vulnerability profiles.

As illustrated in Figure 8, the dasymetric population density surfaces highlight sharp variations in population distribution. Warmer colors (e.g., red) denote densely populated zones concentrated near residential developments, whereas cooler tones (blue and green) indicate lower-density or uninhabited regions dominated by natural land cover.

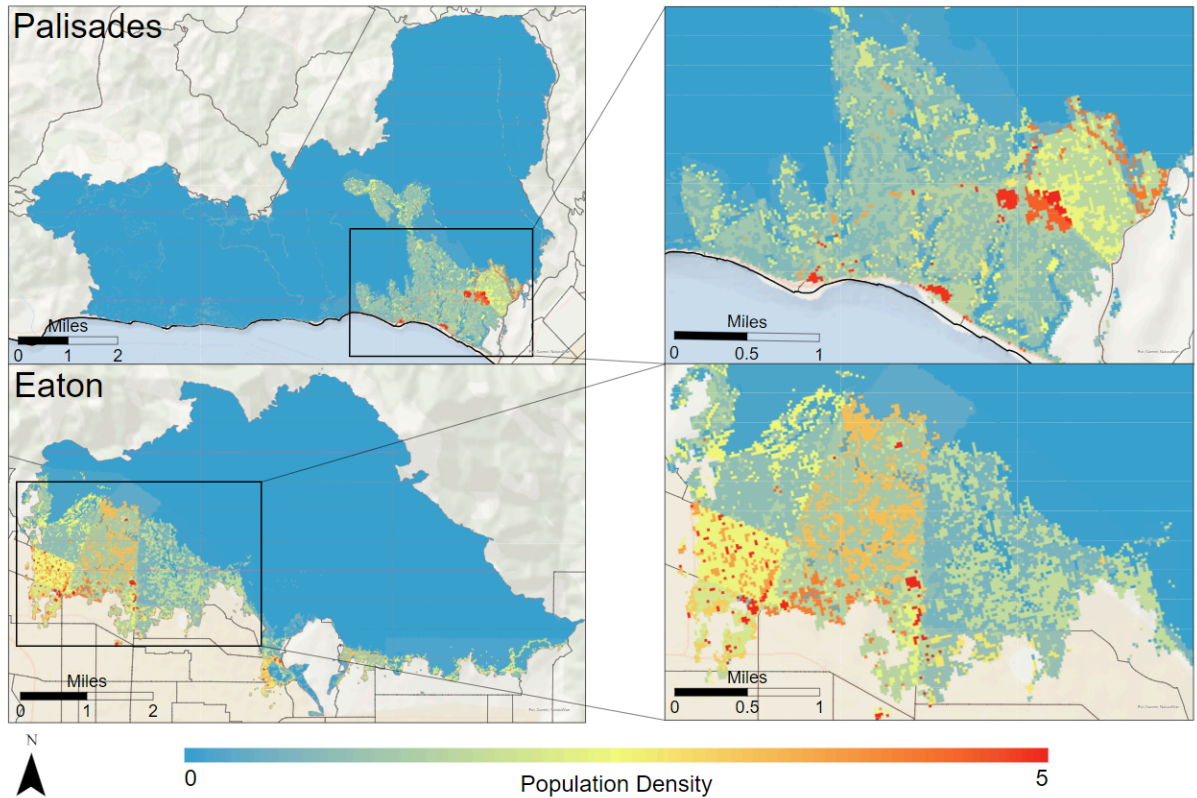

Figure 8. Population density of Palisades and Eaton, Los Angeles by dasymetric mapping

Note: Warmer colors indicate higher population concentrations. Insets highlight detailed variations in population distribution, where high-density zones (in red) cluster near developed areas, and lower-density regions (in blue and green) align with open space and natural land cover.

Figure 9 illustrates daily exposure trends and demographic compositions for Eaton and Palisades, revealing clear spatial and temporal differences. In Palisades, population clusters align with coastal residential areas and developed park-adjacent neighborhoods. In Eaton, the highest population densities are observed along the urban-wildland fringe, particularly near transportation corridors and suburban centers. These patterns validate the effectiveness of dasymetric redistribution in capturing real-world settlement structures and identifying socially vulnerable zones during fire events.

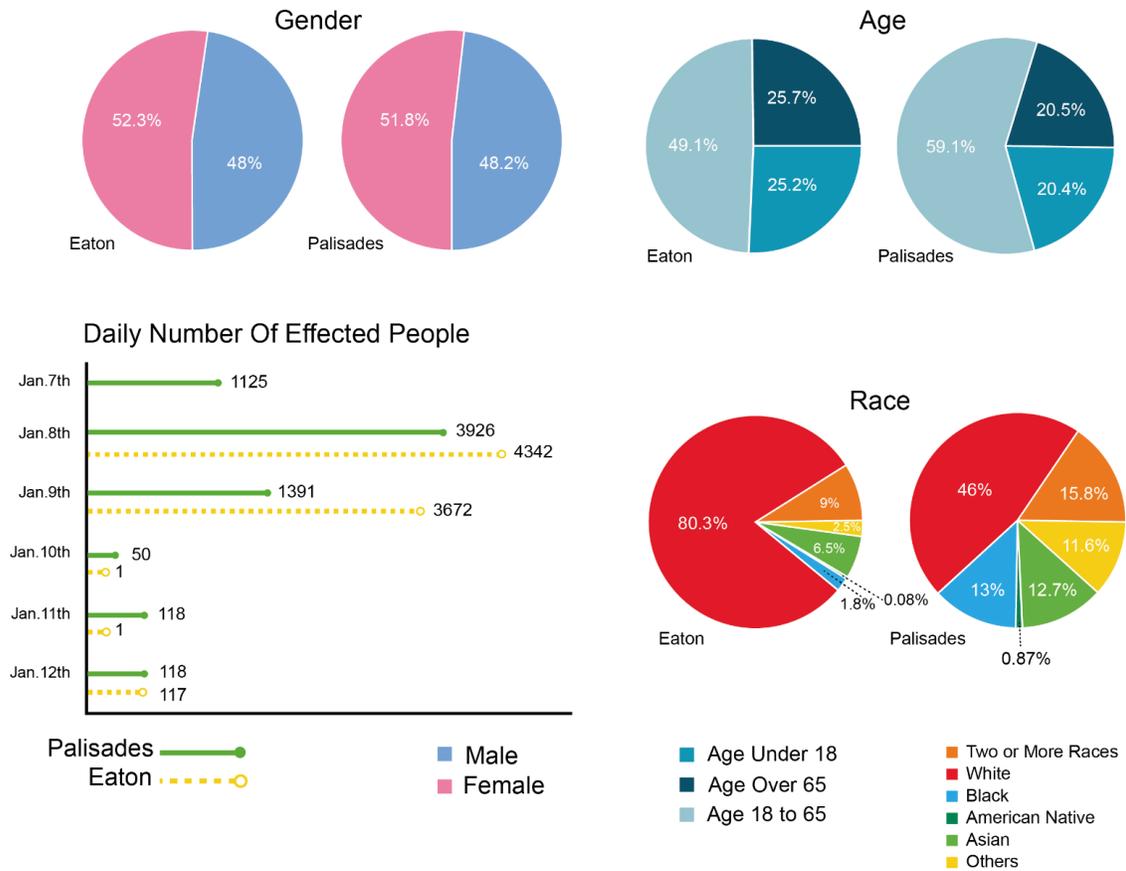

Figure 9. Demographic Characteristics and Daily Population Exposure to Wildfire

Note: Pie charts depict gender, age, and racial composition in Eaton and Palisades. The bar chart shows the number of residents affected daily by wildfire from January 7–12, 2025.

Daily population exposure patterns vary markedly between the two regions. Eaton experiences its peak on January 8, with 4,342 residents exposed—the highest single-day impact across the study period—followed by a sharp decline after January 10, indicating a brief but intense threat. In contrast, Palisades peaks on January 8 and 9 (3,926 and 3,672 residents, respectively), with exposure persisting at low levels through January 12, reflecting a more prolonged and diffuse impact.

Demographic profiles highlight differentiated vulnerabilities. Both areas exhibit near-balanced gender distributions, with a slight female majority (52.3% in Eaton, 51.8% in Palisades), underscoring the need for gender-responsive evacuation and recovery planning. Working-age adults (18–65) constitute the largest group in both regions, but Eaton has a notably higher proportion of seniors (25.7% vs. 20.5%), implying increased mobility and health-related risks. The combined presence of children and elderly in Eaton calls for accessible shelters and medical services. Racial composition further shapes risk profiles. Eaton's affected population is predominantly White (80.3%), while Palisades is more diverse—46% White, 15.8% Asian, 13% multiracial, and smaller shares of other groups—suggesting a greater need for multilingual and culturally adaptive communication strategies.

These social metrics, daily exposure trends, gender balance, age structure, and racial diversity collectively shape each community's resilience and risk profile. The data reinforces the necessity for equity-centered wildfire response planning. For Eaton, concentrated and high-intensity exposure necessitates rapid deployment of resources to senior populations and dense residential zones. In Palisades, extended exposure across a demographically varied population calls for sustained and inclusive community outreach efforts.

*5.4 Comparison of Economic Loss Estimates with Other Assessments*

To contextualize and validate the findings of our tri-environmental framework, we compare our localized economic loss estimates with two prominent regional-scale assessments: the UCLA Anderson Forecast and the Los Angeles County Economic Development Corporation (LAEDC) report (UCLA Anderson, 2025; LAEDC, 2025). While our framework focuses on direct asset-level losses within two specific WUI districts—Palisades and Eaton—the other two studies adopt broader, top-down approaches covering all of Los Angeles County. A structured comparison of the three approaches in terms of estimated losses, scope, methodology, and analytical focus is presented in Table 3.

Table 3. Cross-Comparison of Economic Impact Estimates Assessments by Scope, Methodology, and Focus

| Criteria | Our Study | UCLA Anderson Forecast | LAEDC Report |
| --- | --- | --- | --- |
| Estimated Loss (USD Billion) | 4.855 | 95-164 | 28-53.8 |
| Scope | Direct physical damages to natural, built, and POI assets at WUI level (daily resolution) | County-wide macroeconomic loss including indirect impacts (GDP, capital, employment) | Regional-scale economic disruption including supply chains, jobs, and business continuity |
| Methodology | Daily VIIRS fire data + dasymetric population + land use + POI overlay | Economic modeling using historical wildfire analogs and macroeconomic indicators | Sectoral analysis of economic exposure using regional industry datasets |
| Focus | Asset-level exposure, spatial distribution, daily progression, equity-sensitive diagnostics | Macroeconomic performance, GDP loss, capital depreciation | Business impact, labor markets, regional commerce disruption |

Our analysis estimates direct economic losses of approximately US$ 4.86 billion from the January 2025 wildfire, based on daily fire perimeters and overlays of buildings, roads, and POIs. This estimate, derived from a fine-grained framework integrating satellite fire detections, natural, built, and social environments data, reflects immediate physical damage to natural and built environments over a six-day period (January 7–12).

In contrast, the UCLA Anderson Forecast reports substantially higher losses (US$ 95–164 billion), incorporating indirect and systemic impacts such as GDP contraction, labor disruptions, and capital depreciation, based on macroeconomic simulations rather than spatial data. The LAEDC provides a mid-range estimate (US$ 28–53.8 billion), including sector-specific impacts but aggregated at the county level. While noting 1,117 and 746 disrupted businesses in Palisades and Eaton respectively, it lacks spatial detail on facility type or location.

The differences between our estimate and the two broader assessments arise primarily from three dimensions. First, in the spatial dimension, our study targets neighborhood-level exposure in WUI zones, while the others cover the full urban region. Second, in the dimension of the impact scope, we focus solely on direct damage, whereas the external reports include indirect and long-term consequences. Third, in the methodological dimension, our model processes daily VIIRS fire data with 20-meter spatial resolution, linked to actual infrastructure and demographic distributions, rather than relying on top-down economic projections. Taken together, these distinctions suggest that our framework is particularly well suited for short-term, high-resolution evaluation of direct wildfire impacts, enabling immediate emergency response and equitable resource targeting. In contrast, broader economic assessments are more appropriate for estimating long-term systemic consequences such as GDP loss or labor market disruptions. Although a definitive consensus on the comprehensive economic cost of wildfire remains elusive, our framework provides a distinct, fine-grained perspective grounded in spatially and temporally explicit data, an analytical resolution seldom captured by conventional, top-down economic models.

Despite the disparities in magnitude, the three assessments converge in identifying the wildfire's significant economic footprint and its sectoral impacts. Our high-resolution, tri-environmental approach contributes uniquely to this picture by delivering detailed, temporally dynamic estimates of who and what is affected. This level of granularity supports equitable disaster planning, real-time emergency response, and targeted recovery efforts that would be otherwise masked in regional aggregates.

## 6 . Discussion

### 6.1 Key findings

This study shows that a fine-grained, tri-environmental assessment framework reveals wildfire impacts with temporal and spatial specificity that conventional methods cannot achieve. By integrating daily VIIRS detections, land cover, infrastructure, and dasymetrically mapped population data, the framework captures wildfire dynamics across the natural, built, and social environments. The direct economic loss from the 2025 Los Angeles wildfire totals approximately US$ 4.86 billion, spanning damages to landscapes, infrastructure, and community facilities across the Palisades and Eaton WUI zones.

Palisades experiences its most intense spread and largest losses early (7–9 January) as flames run quickly along coastal ridgelines and into high-value homes and recreation corridors. Eaton, in contrast, sees limited early activity but rapid escalation (9–12 January) as fires descend canyons into dense suburbs. Although residential land dominates the burned area in both zones, Palisades still loses appreciable grassland and open space, signaling greater ecological damage. Built-environment exposure mirrors this timing: Palisades suffers its peak building loss (approximately US$ 1.6 billion) and early road disruption on 7 January, whereas Eaton's building losses surge later (approximately US$ 0.74 billion) alongside heavy damage to neighbourhood service roads. Points-of-interest data add nuance, showing that Eaton's fire front disproportionately affects business, retail, and dining establishments, while Palisades sees broader disruption of community and recreational facilities.

Social impacts follow the same spatial-temporal divide. Eaton's exposure peaks on 8 January, affecting 4,342 residents—older (25.7 % ≥ 65) and mostly White (80 %), posing mobility and outreach challenges. Palisades has two high-exposure days (3,926 and 3,672 on 8–9 January) with more diversity (54 % non-White, larger working-age share), requiring multilingual, sustained relief. These results highlight that wildfire risk in metro WUIs is highly place- and time-specific. A daily, tri-environmental lens is vital to identify who is at risk, what assets are threatened, and when key thresholds are crossed—enabling more targeted, equitable evacuation and recovery.

### *6.2 Contribution to the literature*

This study makes three distinct contributions to wildfire research. First, it operationalizes the concept of a tri-environmental framework—simultaneously evaluating natural, built, and social

dimensions—through a fine-grained, data-integrated approach. Previous applications of this framework in disasters like floods and heatwaves (e.g., Wang et al., 2023; Aquilino et al., 2021) have remained largely theoretical in wildfire science. By integrating daily VIIRS fire detections, high-resolution land-cover maps, infrastructure data from OpenStreetMap, and 20-meter dasymetric population grids, this study translates the tri-environmental concept into a scalable and actionable wildfire analysis tool.

Second, it bridges the gap between high-frequency environmental monitoring and dynamic social vulnerability assessment. By linking daily fire perimeters to disaggregated population surfaces, the study reveals how fire progression affects specific demographic groups—such as children, seniors, or low-income residents—within a census tract. This enables equity-sensitive planning for evacuations, sheltering, and communication strategies, and directly responds to recent calls for disaster research that centers intra-urban diversity and localized needs (Masri et al., 2021).

Third, the study contributes a replicable methodology for wildfire impact modeling that combines daily thermal satellite imagery with KDE-based perimeter estimation, infrastructure overlays, and POI-specific exposure metrics. This integrated pipeline allows for the day-by-day tracking of fire dynamics and their cascading impacts on ecosystems, buildings, and human services. Unlike seasonal or post hoc damage assessments, this workflow provides emergency managers with temporally precise and spatially explicit insights to support real-time decision-making and post-disaster resource allocation.

*6.3 Policy implications*

The results highlight several ways Los Angeles and other wildfire-prone cities can translate a tri-environmental perspective into practice. First, emergency protocols must be timed and tailored to local demographics. Eaton's older population requires evacuation support and medical continuity, while Palisades' cultural diversity demands multilingual alerts and targeted outreach. Land-use policies should reflect localized risk: in Eaton, where business corridors face high losses, fire-resistant standards must apply to shops and mixed-use buildings—not just homes. In Palisades, where public spaces lie within the burn zone, fire-hardening measures like ember-resistant roofs, redundant water, and fire-safe landscaping are essential. Real-time data integration is also key: daily VIIRS detections and short-term spread models should feed into municipal dashboards to enable proactive deployment of personnel, air support, and traffic control.

Second, equity must underpin each intervention. Wildfire exposure in WUI neighborhoods is uneven, and the residents with the least capacity to evacuate, retrofit, or insure their homes often face the highest hazard. To reduce this disparity, cities should expand fuel-management subsidies, low-income insurance incentives, and community-led preparedness programs. However, such strategies only work when vulnerability is defined through a multidimensional lens. By layering demographic characteristics with exposure to natural and built environment threats, cities can more accurately identify the most at-risk streets, facilities, and populations.

Third, the study highlights the importance of integrating high-resolution impact data into long-term resilience planning. Instead of relying solely on post-fire damage surveys or coarse county-level statistics, municipalities should adopt spatially disaggregated assessment tools like the tri-environmental framework used here. This enables planners to evaluate not only where fire has occurred, but also which infrastructure and communities are repeatedly at risk, allowing for more adaptive zoning, strategic retrofits, and targeted investment in both physical and social resilience.

*6.4 Limitations*

This study has several limitations that should be acknowledged. First, the spatial resolution of VIIRS data (375 meters) limits its ability to capture detailed fire behavior at the neighborhood or parcel level. Although kernel density estimation improves the continuity of daily fire perimeters, it cannot fully represent small-scale ignition dynamics or intensity variations within complex urban terrain. The dasymetric mapping approach also assumes static population distribution based on land use, which may not reflect actual daily population shifts due to commuting, tourism, or evacuation. Similarly, our analysis treats all buildings and points of interest equally, without accounting for differences in functional importance, for example, between a home, a business, or a hospital. Exposure is also used as a proxy for impact, which overlooks important mitigating factors such as fire suppression efforts, structure type, or existing defensible space.

In addition, the study is geographically limited to two urban WUI areas in Los Angeles. While Eaton and Palisades provide valuable contrast in topography and demographics, the findings may not fully represent wildfire dynamics or vulnerabilities in other settings. Broader generalization would benefit from applying the framework to a wider range of locations, including rural or lower-income urban peripheries with different infrastructure and social conditions. Future studies could also enhance the model by integrating higher-resolution fire detection, mobility data, and functional weighting of critical infrastructure. Despite these limitations, the tri-environmental framework offers a scalable and replicable structure for advancing urban wildfire impact analysis and informing more targeted and inclusive adaptation strategies.

## 7. Conclusion

This study presents a tri-environmental framework that integrates daily satellite fire detections, natural, built, and social environments data to assess wildfire impacts in Los Angeles's Eaton and Palisades WUI zones. By revealing day-by-day shifts in flame fronts, quantifying direct losses to land, roads, buildings, and community services, and pinpointing which age, income, and racial groups are placed at greatest risk, the framework demonstrates that wildfire is simultaneously an ecological, infrastructural, and social crisis whose effects vary block by block. These results show that generic, city-wide strategies are inadequate; effective planning must match interventions—evacuation support, land-use zoning, fuel management, and equitable recovery funds—to the unique natural, built, and demographic profiles of each neighbourhood. Because the workflow relies on globally available satellite feeds and open data, it can be readily replicable to other WUI regions and expanded with real-time mobility streams, critical-infrastructure weights, or machine-learning spread models. As climate change and urban expansion intensify wildfire threats, such integrated, high-resolution approaches will be essential for designing communities that are not only fire-adapted but also socially just.

# Appendix A. Appendix

Table A1.1 Daily Impacts of the 2025 Los Angeles Wildfire on Natural and Built Environments

| District | Category | Unit | Jan.7th | Jan.8th | Jan.9th | Jan.10th | Jan.11th | Jan.12th |
|---|---|---|---|---|---|---|---|---|
| Palisades | Natural Environment | US Dallor | 1561390391 | 1240871065 | 1780424622 | 239104712 | 4185981.21 | 10129179.83 |
| Palisades | Built Environment: Buildings | US Dallor | 15947987930 | 6593928864 | 40966682.04 | 20483341.02 | 25604176.28 | 4702633710 |
| Palisades | Built Environment: Roads | US Dallor | 890838.1864 | 236187.7785 | 100049.2768 | 46145.95487 | 45256.20527 | 96315.70797 |
| Palisades | Built Environment: Points of Interest | Number | 220 | 1397 | 1045 | 14 | 6 | 8 |
| Eaton | Natural Environment | US Dallor | - | 361019911.5 | 398404510.6 | 11726789.54 | 13995961.07 | 196306467.1 |
| Eaton | Built Environment: Buildings | US Dallor | - | 7910281780 | 3565599.18 | 6239798.56 | 106967975.4 | 7369202104 |
| Eaton | Built Environment: Roads | US Dallor | - | 891744.5029 | 62707.70267 | 47445.12271 | 31006.23706 | 1068464.71 |
| Eaton | Built Environment: Points of Interest | Number | - | 901 | 915 | 2 | 0 | 6 |

Table A1.2 Composition of Fire-Affected Land Cover Types in the Natural Environment

| Category: Natural Environment | Palisades | Eaton |
|---|---|---|
| Residential | 92.73% | 98.32% |
| Industrial | 4.43% | 0.26% |
| Grass | 2.69% | - |
| Others | 0.15% | 1.42% |

Table A1.3 Composition of Fire-Affected Roads in the Built Environment

| Category: Roads | Palisades | Eaton |
|---|---|---|
| Pedestrian Paths | 42.76% | 10.21% |
| Primary Roads | 28.45% | 30.06% |
| Residential & Service Roads | 6.21% | 78.91% |
| Tracks | 2.47% | 9.15% |
| Unclassified Roads | 0.01% | 0.00% |

Table A1.4 Composition of Fire-Affected Points of interst in the Built Environment

| Category: Points of Interest | Palisades | Eaton |
|---|---|---|
| Business and Prof. Services | 24.08% | 34.74% |
| Community and Government | 5.23% | 13.55% |
| Dining and Drinking | 9.94% | 9.41% |
| Landmarks and Outdoors | 6.85% | 14.99% |
| Retail | 11.46% | 13.63% |
| Unknown | 6.28% | 7.05% |
| Others | 12.76% | 13.64% |